\documentclass{aa}  

\usepackage{txfonts} 
\usepackage{natbib}
\usepackage{graphics}
\usepackage{graphicx}
\usepackage{amssymb}
\usepackage{multicol}
\bibpunct{(}{)}{;}{a}{}{,}
\def\Teff{\ensuremath{T_{\mathrm{eff}}}}

\def\logg{\ensuremath{\log g}}
\def\vsini{\ensuremath{{\upsilon}\sin i}}
\def\kms{$\mathrm{km\,s}^{-1}$}

\begin{document}

\title{Search for stellar spots in field blue horizontal-branch stars}

\author{E.~Paunzen\inst{1} 
\and K.~Bernhard\inst{2,3}
\and S.~H{\"u}mmerich\inst{2,3}
\and J.~Jan{\'i}k\inst{1}
\and E.~A.~Semenko\inst{4}
\and I.~A.~Yakunin\inst{4}}
\institute{Department of Theoretical Physics and Astrophysics, Masaryk University,
Kotl\'a\v{r}sk\'a 2, CZ-611\,37 Brno, Czech Republic
\email{epaunzen@physics.muni.cz}
\and{American Association of Variable Star Observers (AAVSO), 49 Bay State Rd, Cambridge, MA, 02138, USA}
\and{Bundesdeutsche Arbeitsgemeinschaft für Ver{\"a}nderliche Sterne e.V. (BAV), D-12169, Berlin, Germany}
\and{Special Astrophysical Observatory of the Russian Academy of Sciences, Nizhnii Arkhyz 369167, Russia}
}

\date{} 
\abstract
{Blue horizontal-branch stars are Population II objects which are burning helium in their core and possess a 
hydrogen-burning shell and radiative envelope. Because of their low rotational velocities, diffusion has been predicted to work in their atmospheres. In many respects, blue horizontal-branch stars closely resemble the magnetic chemically peculiar stars of the upper main sequence, which show photometric variability caused by abundance spots on their surfaces. These spots are thought to be caused by diffusion and the presence of a stable magnetic field. However, the latter does not seem to be axiomatic.}
{We searched for rotationally induced variability in 30 well-established bright field blue horizontal-branch stars in the solar neighbourhood and searched the literature for magnetic fields measurements of our targets.}
{We employed archival photometric time series data from the All Sky Automated Survey (ASAS), All-Sky Automated 
Survey for Supernovae (ASAS-SN), and Wide Angle Search for Planets (SuperWASP) surveys. The data were carefully 
reduced and processed, and a time series analysis was applied using several different techniques. We also synthesized
existing photometric and spectroscopic data of magnetic chemically peculiar stars in order to study possible different surface 
characteristics producing lower amplitudes.}
{In the accuracy limit of the employed data, no significant variability signals were found in our sample stars. 
The resulting upper limits for variability are given.}  
{We conclude that either no stellar surface spots are present in field blue horizontal-branch stars, or their characteristics (contrast, 
total area, and involved elements) are not sufficient to produce amplitudes larger than a few millimagnitudes in the optical 
wavelength region. New detailed models taking into account the elemental abundance pattern of blue horizontal-branch stars are needed to synthesize 
light curves for a comparison with our results.}


\titlerunning{}
\authorrunning{}
\maketitle

\section{Introduction} \label{introduction}

Blue horizontal-branch (BHB) stars were first unambiguously detected and defined in Galactic globular clusters \citep{Arp52}. These stars burn helium in their core and possess a hydrogen-burning shell and a radiative envelope. Owing to their complexity, 
some aspects of the formation and internal structure of horizontal-branch stars, in general, have been a matter of 
debate for decades \citep{Catel09}. For instance, the blue extension of the horizontal branch varies among clusters, a fact partly 
associated with metallicity but not entirely explained by it. A number of secondary parameters, such as stellar rotation, cluster concentration, presence of super-oxygen-poor stars, cluster mass, 
environment of formation, and cluster age have been suggested to provide an explanation for this behaviour, but 
none of these have been proven completely adequate in describing the complex observational picture \citep{Milon14}. 

The field horizontal-branch (FHB) stars in the solar neighbourhood have been detected in recent decades mainly owing to
their kinematic characteristics as they are ideal tracers of the kinematics and dynamics of the
thick disk throughout the inner halo of the Milky Way \citep{Wilhe99}.

\begin{table*}[t]
\caption{Astrophysical parameters and their errors, taken from \citet{Kinm00} and \citet{Behr03} according to the
column ``Ref''. The [M/H] values are according to \citet{Kinm00} whereas [Fe/H] is taken from \citet{Behr03}.
The $V$ magnitudes are from \citet{Kharc01}.}
\label{stars_parameters}
\begin{center}
\begin{tabular}{lcccccccc}
\hline
Star & TYC & $V$ & \Teff & \logg & [M/H] & [Fe/H] & \vsini & Ref \\
& & (mag) & (K) & & (dex) & (dex) & (\kms) & \\
\hline
BD$-$07 230     & 4684--1840--1 & 11.13 & $9\,647\pm269$        & $3.40\pm0.33$ & & $-0.44\pm0.11$ & 2 & 2 \\
BD+01 548       & 55--357--1    & 10.76 & $8\,714\pm198$        & $3.38\pm0.16$ & & $-2.23\pm0.06$ & 10 & 2 \\
BD+25 2602      & 1994--1778--1 & 10.10 & $8\,410\pm100$        & $3.17\pm0.04$ & --2.0   & $-2.08\pm0.11$ & 12 & 1 \\
BD+42 2309      & 3020--678--1  & 10.84 & $8\,825\pm88$ & $3.20\pm0.05$ & --1.5   & $-1.69\pm0.05$ & 35 & 1 \\
HD 2857 & 4678--90--1   & 9.90  & $7\,566\pm58$ & $3.00\pm0.08$ & --1.5 & $-1.67\pm0.11$ & 30 & 1 \\
HD 4580 & 3659--586--1  & 8.78  & $8\,453\pm80$ & $3.20\pm0.08$ & --1.5 & & 14 & 1 \\
HD 8376 & 2292--1267--1 & 9.60  & $8\,133\pm48$ & $3.27\pm0.04$ & --2.5 & & 10 & 1 \\
HD 13780        & 8045--420--1  & 9.82  & $7\,930\pm23$ & $3.12\pm0.02$ & --1.5   & & 14 & 1 \\
HD 14829        & 5282--1684--1 & 10.30 & $9\,086\pm245$        & $3.31\pm0.19$ & & $-2.01\pm0.31$ & 14 & 2 \\
HD 31943        & 7593--771--1  & 8.26  & $7\,893\pm59$ & $3.22\pm0.11$ & --1.0   & & 7 & 1 \\
HD 60778        & 4830--1230--1 & 9.09  & $8\,072\pm44$ & $3.13\pm0.03$ & --1.5   & $-1.48\pm0.06$ & 11 & 1 \\
HD 74721        & 816--2644--1  & 8.70  & $8\,908\pm88$ & $3.31\pm0.02$ & --1.5   & $-1.41\pm0.04$ & 2 & 1 \\
HD 78913        & 9195--1020--1 & 9.28  & $8\,515\pm355$        & $3.25\pm0.01$ & --1.5   & & 14 & 1  \\
HD 86986        & 835--1408--1  & 7.98  & $7\,936\pm47$ & $3.19\pm0.03$ & --1.5   & $-1.85\pm0.05$ & 4 & 1 \\
HD 87047        & 2503--1098--1 & 9.72  & $7\,828\pm49$ & $3.07\pm0.04$ & --2.5   & $-2.36\pm0.10$ & 0 & 1 \\
HD 87112        & 3821--809--1  & 9.70  & $9\,733\pm38$ & $3.48\pm0.01$ &       --1.5   & $-1.65\pm0.07$ & 3 & 1 \\
HD 93329        & 849--785--1   & 8.77  & $8\,237\pm82$ & $3.12\pm0.05$ & --1.5   & $-1.49\pm0.08$ & 7 & 1 \\
HD 106304       & 7760--1284--1 & 9.06  & $9\,747\pm365$        & $3.50\pm0.03$ & --1.5   & & 10 & 1 \\
HD 109995       & 3018--494--1  & 7.59  & $8\,514\pm123$        & $3.08\pm0.07$ & --1.5 & $-1.76\pm0.09$ & 25 & 1 \\
HD 117880       & 6125--410--1  & 9.06  & $9\,285\pm131$        & $3.34\pm0.09$ & --1.5   & $-2.25\pm0.14$ & 13 & 1 \\
HD 128801       & 911--1261--1  & 8.73  & $10\,313\pm163$       & $3.55\pm0.02$ & --1.5   & $-1.38\pm0.09$ & 4 & 1 \\
HD 130095       & 6754--156--1  & 8.14  & $9\,010\pm90$ & $3.30\pm0.03$ & --2.0 & & 5 & 1 \\
HD 130201       & 8279--2--1    & 10.06 & $8\,645\pm275$        & $3.48\pm0.03$ & --1.5   & & 16 & 1 \\
HD 139961       & 7849--447--1  & 8.84  & $8\,517\pm83$ & $3.23\pm0.07$ & --1.5   & & 39 & 1 \\
HD 161817       & 2081--3673--1 & 6.98  & $7\,533\pm10$ & $3.00\pm0.01$ & --1.5   & $-1.52\pm0.05$ & 17 & 1 \\
HD 167105       & 3536--45--1   & 8.94  & $9\,025\pm111$        & $3.29\pm0.04$ & --1.5   & $-1.62\pm0.08$ & 21 & 1 \\
HD 180903       & 6879--187--1  & 9.57  & $7\,683\pm44$ & $3.10\pm0.01$ & --1.5   & $-1.73\pm0.08$ & 17 & 1 \\
HD 213468       & 8003--27--1   & 10.81 & $9\,147\pm68$ & $3.28\pm0.02$ & --1.5   & & 12 & 1 \\  
HD 252940       & 1885--326--1  & 9.08  & $7\,563\pm37$ & $2.95\pm0.05$ & --1.5   & $-1.70\pm0.09$ & 24 & 1 \\
HZ 27   &       & 12.81 & $9\,883\pm289$        & $3.38\pm0.49$ & & $-1.39\pm0.35$ & 7 & 2 \\  
\hline  
\multicolumn{8}{l}{1: \citet{Kinm00}, 2: \citet{Behr03}} \\                
\end{tabular}    
\end{center}                                      
\end{table*}

\begin{table*}[t]
\caption{Characteristics of the employed time series (time baseline and number of observations) and corresponding upper limits of variability.}
\label{stars_tsa}
\begin{center}
\begin{tabular}{lccccccccc}
\hline
& \multicolumn{3}{c}{ASAS} & \multicolumn{3}{c}{ASAS-SN} & \multicolumn{3}{c}{SuperWASP} \\
Star & $\Delta t$ & $N$ & UL & $\Delta t$ & $N$ & UL & $\Delta t$ & $N$ & UL \\
& (d) & & (mmag) & (d) & & (mmag) & (d) & & (mmag) \\
\hline
BD$-$07 230     & 3278.1        & 378 &         7.2 &   2189.8  & 970   & 1.7 & & & \\
BD+01 548       & 3232.1        & 359   & 5.3   & 1963.6 &      902     & 2.9 & & & \\
BD+25 2602      &       2267.9  &       206     &       6.2     &       2225.8  &       799     &       1.9     &       1123.1  &       7201    &       2.4     \\
BD+42 2309      &               &               &               &       2250.7  &       649     &       2.9     &       1476.0  &       5664    &       1.6     \\
HD 2857 &       3296.0  &       410     &       4.5     &       2173.9  &       928     &       3.4     &               &               &               \\
HD 4580 &               &               &               &       1107.9  &       887     &       10.3    &       765.1   &       1425    &       1.4     \\
HD 8376 &               &               &               &       2150.9  &       1284    &       1.2     &       1501.0  &       7511    &       1.0     \\
HD 13780        &       3256.1  &       497     &       4.0     &       1375.6  &       768     &       5.2     &               &               &               \\
HD 14829 & 3296.0       & 469   & 4.7   & 2192.7        & 1007 &        1.8 &   &   & \\
HD 31943        &       3300.0  &       1013    &       2.8     &       1400.0  &       820     &       11.1    &       151.9   &       557     &       1.4     \\
HD 60778        &       3269.1  &       458     &       3.1     &       2230.9  &       1008    &       3.0     &               &               &               \\
HD 74721        &       2543.1  &       399     &       3.9     &       2217.8  &       651     &       11.8    &               &               &               \\
HD 78913        &       3261.0  &       1523    &       4.0     &       1403.3  &       961     &       4.7     &               &               &               \\
HD 86986        &       2366.7  &       301     &       6.5     &       2155.7  &       667     &       10.3    &               &               &               \\
HD 87112        &               &               &               &       2203.9  &       752     &       3.8     &               &               &               \\
HD 87047        &               &               &               &       2214.0  &       1243    &       1.4     &       777.0   &       3623    &       0.9     \\
HD 93329        &       2864.1  &       403     &       4.2     &       2171.8  &       722     &       6.1     &               &               &               \\
HD 106304       &       3295.0  &       645     &       2.8     &       1391.0  &       815     &       10.0    &       754.2   &       15303   &       1.0     \\
HD 109995       &               &               &               &       2212.0  &       676     &       10.8    &               &               &               \\
HD 117880       &       3198.6  &       657     &       4.1     &       2203.8  &       648     &       13.6    &               &               &               \\
HD 128801       &       2924.1  &       402     &       4.5     &       2218.8  &       749     &       7.8     &               &               &               \\
HD 130095       &       3170.7  &       470     &       4.3     &       2162.7  &       766     &       9.9     &       752.3   &       4192    &       2.3     \\
HD 130201       &       3190.7  &       540     &       3.4     &       731.1   &       565     &       2.9     &       752.3   &       7082    &       1.1     \\
HD 139961       &       3189.6  &       648     &       4.2     &       731.1   &       376     &       18.7    &               &               &               \\
HD 161817       &       2360.6  &       270     &       4.5     &       1996.3  &       859     &       26.3    &               &               &               \\
HD 167105       &               &               &               &       2121.1  &       784     &       10.9    &       1548.0  &       7406    &       0.9     \\
HD 180903       &       3129.7  &       483     &       3.5     &       1452.8  &       727     &       4.2     &       698.3   &       8404    &       1.6     \\
HD 213468       &       3259.1  &       423     &       4.6     &       1333.7  &       774     &       2.8     &       554.7   &       8117    &       1.0     \\
HD 252940       &       2543.1  &       264     &       5.1     &       1179.7  &       435     &       10.4    &       918.8   &       2576    &       1.1     \\   
HZ 27 & & & & 2223.8 &  725     & 2.9   & 1476.0        & 9176  & 1.6 \\              
\hline   
\end{tabular}    
\end{center}                                      
\end{table*}

An intriguing phenomenon was first reported by \citet{Behr99}, who found large deviations in element 
abundances from the expected cluster metallicity for BHB stars in the globular cluster NGC~6205. For 
example, iron was found to be a factor of three enhanced compared to the solar value, or about 100 
times the mean cluster iron abundance. Such atmospheric effects are well known and studied in the 
classical chemically peculiar (CP) stars of the upper main sequence \citep{Preston74}, which share 
the same temperature regime as the BHB stars. Later on, \citet{Khala10} found clear evidence of 
vertical stratification of Fe in atmospheres of BHB stars. In addition, it was reported that BHB stars with $T_\mathrm{eff}$\,$\geq$\,11\,500\,K have lower rotational velocities than their cooler 
counterparts \citep{Behr03}, suggesting that the atmospheres of such stars are stable enough for atomic diffusion to work. The models by \citet{Quiev09} showed that He sinks in stars with low rotational 
velocities, which leads to the disappearance of the superficial He convection zone. This then 
opens the door for atomic diffusion to play a role. The atmospheric models of \citet{Hui00} 
showed that the observed photometric jumps and gaps for hot BHB stars can be explained by elemental 
diffusion in their atmospheres. These models self-consistently calculate the structure of the atmosphere 
while taking into account the stratification predicted by diffusion (assuming equilibrium). They confirm 
that vertical stratification of the elements can strongly modify the structure of the atmospheres of BHB
stars \citep{LeBlanc09}. Such structural changes of the atmosphere lead to the photometric anomalies discussed above.

For the magnetic CP (mCP) stars, the overabundant elements in their atmospheres are concentrated into large 
spot regions that persist for decades to centuries. As an mCP star rotates, periodic variations in its brightness, 
spectrum, and magnetic field are observed, in which case the star is referred to as an $\alpha^2$ Canum Venaticorum 
(ACV) variable. These variations led to the development of the oblique rotator model \citep{Stibbs50}, which explains 
the periodicity of the variations as a geometrical effect as the star rotates, using a simple dipolar geometry. 
However, the search for strong magnetic fields in FHB stars was not successful \citep{Elkin98}.

In this work, we recall the case of the HgMn (CP3) stars, which do not show strong large-scale organized magnetic fields 
\citep{Kochu13}. However, the line-profile variations detected in the spectra of these stars have also been interpreted 
in the terms of abundance inhomogeneities \citep{Hubrig06}. Therefore, rotationally induced photometric variability at 
some level would be expected. While photometric variations in CP3 stars have been established beyond doubt, the underlying 
mechanism is still a matter of debate \citep{Morel14}. However, recent studies strongly favour rotational modulation over 
pulsation \citep[cf. the discussion in][]{Huemm18}.

Recently, \citet{Balona17} summarized evidence of the presence of rotational modulation in the \textit{Kepler} light curves 
of mid A- to late B-type stars. He concluded that the time-frequency diagrams show stochastic variations in all respects similar 
to those in spotted cool stars. Following this interpretation, more than half of his sample of stars are proposed to show rotational 
modulation, indicating that starspots may be the rule rather than the exception among A-type stars. However, these findings still 
need to be verified with independent photometric time series, and, in particular, spectroscopic data.

Owing to the striking similarities of slowly rotating BHB and classical CP stars, we selected the brightest 
FHB stars in the solar neighbourhood to search for spot-induced variability. To our knowledge, such an analysis has never been attempted before. Looking 
at the BHB stars, we also find RR Lyrae variables \citep{For11} and low-amplitude $g$-mode pulsators \citep{Osten12}. The latter are 
technically hot subdwarfs of the sdB type, as they appear as B-type stars with a rather high surface gravity; this places these pulsators below the 
main sequence in the Hertzsprung–Russell diagram. However, their surface gravities are lower and their helium abundances much higher 
than in the majority of sdB stars, making them quite unusual.

In total, we present a detailed time series analysis of 30 targets using All Sky Automated Survey (ASAS), 
All-Sky Automated Survey for Supernovae (ASAS-SN), and Wide Angle Search for Planets (SuperWASP) data. The results and astrophysical 
characteristics of our sample stars are compared to CP stars.

\section{Target selection, magnetic field measurements, and photometric data treatment} \label{target_data}

This section provides information about the employed photometric time series data and describes the processes of target selection, data reduction, and data analysis.

\subsection{Target selection}

We have used the works of \citet{Kinm00}, \citet{Behr03}, and \citet{Kafan16} as a basis for our investigation. These authors analysed bright FHB stars in the solar neighbourhood and derived astrophysical parameters and elemental abundances. From their list, we excluded RR Lyrae stars because the large amplitude pulsations of these objects interfere with the diffusion mechanism, thereby effectively preventing the formation of stellar spots.

For the remaining stars, we searched for time series in the ASAS, ASAS-SN, and SuperWASP archives. The data from ASAS and SuperWASP were already successfully applied in the analysis of ACV variables \citep{Bernh15a,Huemm16}. In total, 30 stars fulfilled these criteria. Table \ref{stars_parameters} lists the astrophysical parameters of our target stars. None of these objects have data in the CoRoT \citep{DeMed13} and Kepler/K2 \citep{Koch10,Lund16} photometric databases.

\subsection{Magnetic field measurements} \label{magnetic_field_literature}

\citet{Elkin98} presented magnetic field measurements of eight FHB stars, including HD 60778, HD 74721, HD 86986, and HD 161817 from our sample. He used classical Zeeman spectroscopy which was also successfully applied to mCP stars \citep{Bychk09}. The magnetic field measurements have been based on circularly polarized spectra obtained with the 6\,m telescope of the Russian Academy of Sciences Special Astrophysical Observatory. The field strength (or corresponding upper limit) was inferred from the relative Zeeman displacements of the spectral lines recorded in the right and left circular polarizations.

For HD 60778, HD 74721, and HD 86986, the following measurements were published: $+150\pm115$\,G, $+240\pm150$\,G, and $-430\pm580$\,G, respectively. For HD 161817, four individual measurements are available: $-550\pm140$\,G, $-90\pm80$\,G, $+30\pm110$\,G, and $+100\pm160$\,G. These results seem to indicate the possible presence of a stable magnetic field in HD 161817, which varies over the rotational period. Unfortunately, these are the only measurements available for this star in the literature. Clearly, new spectropolarimetric observations are needed to shed more light on this important issue, although we caution that -- as stated above (cf. Section \ref{introduction}) -- it is still a matter of debate whether a stable magnetic field is a necessary requirement for the formation of stellar spots.

\subsection{Data sources, reduction, and analysis} \label{data_sources}

The ASAS project aims at continuous photometric monitoring of the whole sky and has the ultimate goal of detecting and 
investigating any kind of photometric variability. We employed data from the third phase of the project, ASAS-3. The typical exposure time for ASAS-3 $V$-filter observations is three 
minutes, which results in reasonable photometry for stars in the magnitude range 7\,$<$\,$V$\,$<$\,14\,mag. In general, a 
field is observed one to three times per day \citep{Pigul14}.

The ASAS-SN survey has been imaging the entire visible sky every night to a depth of $V$\,$<$\,17\,mag \citep{Kocha17}. The available 
data span up to five years of observations. As of end of 2017, ASAS-SN comprises five stations each consisting of four 14\,cm aperture 
Nikon telephoto lenses. Observations are made using $V$ (two stations) or $g$ (three stations) band filters and three dithered 90\,s 
exposures. ASAS-SN saturates at 10 to 11\,mag, where the exact limit depends on the camera and the image position. However, a procedure 
inherited from the ASAS survey is applied which corrects for saturation but introduces a larger noise in the corresponding data sets \citep{Pojma02}.

The SuperWASP survey started in 2004 and covers both hemispheres. It provides long-term photometric time series in a broadband filter 
(4\,000 -- 7\,000\,\AA) with an accuracy better than 1\% for objects in the magnitude range 8\,$<$\,$V$\,$<$\,11.5\,mag \citep{Polla06}. 
Observations consist in general of two consecutive 30\,s integrations followed by a 10-minute gap. In this work, we use data from the first 
and only WASP public data release \citep{Butte10}\footnote{https://wasp.cerit-sc.cz/}.

Data from the ASAS and SuperWASP surveys have been successfully employed to investigate the small amplitude photometric variability of mCP 
stars \citep{Bernh15a,Bernh15b,Huemm16}. These data should therefore be well suited to investigate the possible occurrence of spots in our 
target stars, which we expect to exhibit similar amplitudes. As yet, ASAS-SN data have not been used for the identification of ACV variables. 
Before we applied a comprehensive time series analysis, different data reduction steps for the corresponding surveys were performed, which 
are described in more detail below.

Measurements with quality assignments ``C'' and ``D'' were excluded from the ASAS data. The mean $V$ magnitudes were calculated as the 
weighted average of the values provided in the five different apertures indicated by the ASAS system \citep{Kovac05}. To check the 
feasibility of this approach, we subsequently restricted our analysis to the ``best'' aperture, as indicated by the ASAS system for a 
given star. No significant differences were found between the two approaches. Finally, a basic 5$\sigma$ clipping was performed to clean 
the light curves from outliers.

ASAS-SN measurements are taken with different cameras, which we treated separately. The mean for each individual data set was calculated 
and data points were deleted on a 5$\sigma$ basis. After that, the data of the individual cameras were merged.

We excluded measurements with an error larger than 0.05\,mag from the SuperWASP data sets. Each camera was treated separately; the mean 
for each individual data set was derived, and data points were deleted on a 5$\sigma$ basis. This procedure also corrects for the different 
offsets of the cameras. As last step, the data of the individual cameras were merged. However, for larger data sets, we also investigated 
the measurements from each camera separately.

The resulting light curves were examined in more detail using the programme package {\sc Period04} \citep{Lenz05}, which performs a discrete Fourier transformation. The shortest expected periods are about 0.5\,d, which was consequently chosen as limit for the investigated period range in the analysis of all data sets. For rotational velocities larger than that, meridional circulation effectively prevents diffusion in stellar atmospheres \citep{Charb91}.

The results from {\sc Period04} were checked with {\sc cleanest} and phase dispersion minimization (PDM) algorithms as implemented in the programme package {\sc Peranso} \citep{Paunz16}. The same results were obtained within the derived errors, which depend on the time series characteristics, i.e. the distribution of the measurements over time and photon noise.

\begin{figure}
\begin{center}
\includegraphics[width=80mm, clip]{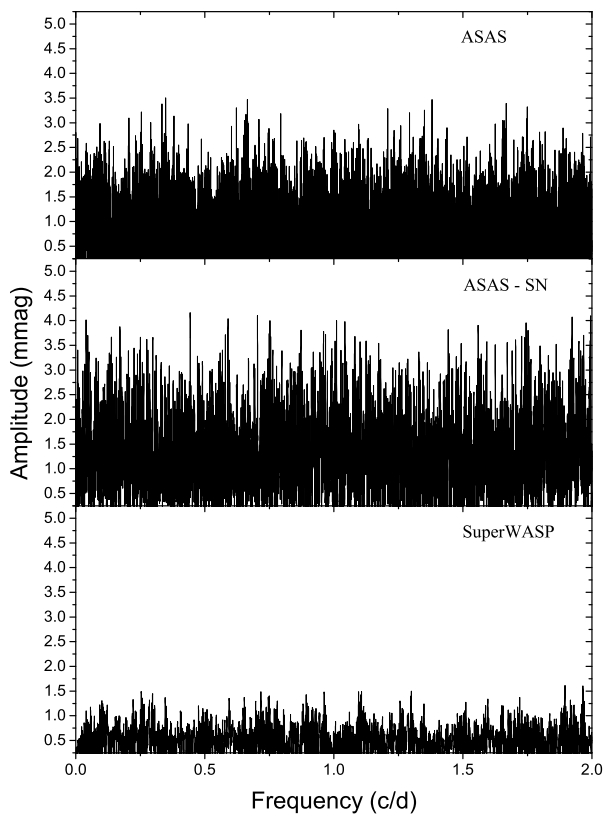}
\caption{Amplitude spectra of the ASAS, ASAS-SN, and SuperWASP data for HD 180903. We define the upper limit of variability 
(``UL'' in Table \ref{stars_tsa}) as upper envelope of the peaks in an amplitude spectrum, which is a very conservative definition.}
\label{hd180903_as} 
\end{center} 
\end{figure}

Defining the upper limit of variability is not straightforward and has often been discussed in the literature \citep{Reegen08}. In general, 
the statistical significance of the noise in the Fourier spectrum is underestimated. We employ a conservative approach and define the 
upper limit of variability as the upper envelope of the peaks in an amplitude spectrum (Fig. \ref{hd180903_as}). Table \ref{stars_tsa} gives the characteristics of the employed photometric time series.

\section{Discussion} \label{analysis}

The investigated FHB stars share common characteristics with classical CP stars. While these objects are found between spectral types early B to early F, our sample stars cover the \Teff\ range from about 10\,500 to 7\,500\,K, which corresponds to spectral types B9 to A8 on the main sequence. Thus, neither convection nor stellar winds should significantly affect their atmospheres. In addition, the observed \vsini\ values (cf. Table \ref{stars_parameters}) of the FHB stars are well in the range of the values for CP stars. Therefore, in principle, FHB stars should fulfil the requirements for efficient diffusion in their stellar atmospheres, which is a necessary condition for the formation of atmospheric abundance spots. Furthermore, available magnetic field measurements point to the possible presence of a stable magnetic field in one of our sample stars (HD 161817; cf. Section \ref{magnetic_field_literature}), although more observations are needed to confirm this. However, we caution that the complete framework and setting in which the ``CP phenomenon'' occurs is still not well understood, which also holds true for the prerequisites of efficient diffusion in their stellar atmospheres and the timescales involved \citep{Stift16}.

For the investigation of the presence of spot-induced variability in our sample of FHB stars, a total of 66 individual data sets with about 125\,000 individual measurements was analysed. Overlapping data sets from all three surveys exist for eight stars (BD+25 2602, HD 31943, HD 106304, HD 130095, HD 130201, HD 180903, HD 213468, and HD 252940). Time bases range from half a year to almost ten years, which is much longer than the expected rotational periods of up to a few days. With our sample of 30 stars, we can exclude that observational effects such as inclination or rotational periods much longer than the time basis of the photometric data impact our results in a significant way. Thus, given the known \vsini\ values and assuming randomly distributed inclination angles \citep{Netop17}, rotationally induced variability should be readily detectable with the employed data in this study.

However, we were not able to detect variability with a statistically significant amplitude for any of our sample stars. Table \ref{stars_tsa} lists the derived upper limits of variability as defined in Sect. \ref{target_data}. As an example, Fig. \ref{hd180903_as} presents the amplitude spectra for all analysed data sets of HD 180903. Similar results have been obtained for the other targets.

\begin{figure}[t]
\begin{center}
\includegraphics[width=80mm,clip]{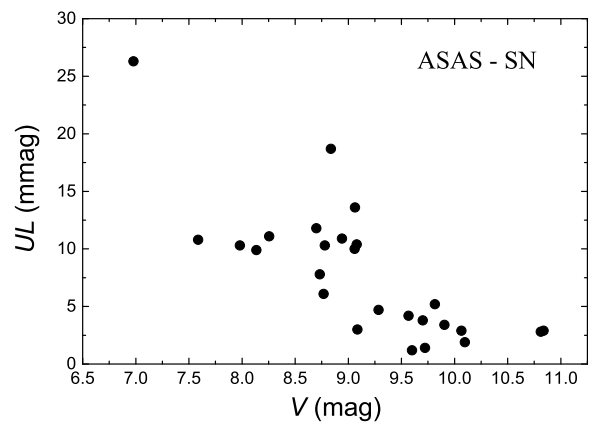}
\caption{Dependence of the upper limit for variability (UL) on the visual magnitude of the target stars from ASAS-SN data. Owing to 
non-linearity effects, the values for stars brighter than 9th magnitude are much larger than for the fainter targets.}
\label{v_ul} 
\end{center} 
\end{figure}

In general, the upper limits derived from SuperWASP data (about 1.5\,mmag) are a factor two lower than the limits derived from ASAS, which is because of the much denser time coverage of SuperWASP. For the ASAS-SN data, we found an obvious correlation between the derived upper limits and the visual magnitudes of the corresponding target stars (Fig. \ref{v_ul}). For stars brighter than the ninth magnitude, the upper limits increase up to 25\,mmag, which is likely due to the treatment of saturated pixels as described in \citet{Kocha17}. Nevertheless, for the fainter targets, the upper limits derived from ASAS-SN are in line with the values from ASAS and SuperWASP data (Table \ref{stars_tsa}).

ASAS and SuperWASP data have been successfully employed to investigate the spot-induced light changes of ACV variables \citep{Bernh15a,Bernh15b,Huemm16}. We compared the derived upper limits for our FHB stars to the results for ACV variables published by these investigators. As the ACV stars are also within the same magnitude range as our sample stars, the quality of the employed time series data is comparable, which renders a direct comparison possible. However, it is important to note that the ACV samples are significantly dominated by A0 Si stars (\Teff\ of about 10\,000\,K), which are known to exhibit the largest photometric amplitudes among mCP stars.

ACV variables exhibit variability amplitudes from 3 to 35\,mmag in ASAS data (mean value of 10\,mmag) and 2.8 to 53\,mmag (mean value of 11\,mmag) in SuperWASP data. Unfortunately, no information about the non-variable mCP stars and the corresponding upper limits are given in the above-listed references. However, the lowest detected amplitudes are about 2\,mmag and the noise should be at least a factor of two less \citep{Huemm16}. Therefore, typical ACV-like variations would have been readily detectable in most of the data sets we employed for the analysis of our sample FHB stars, which suggests that either no spots are present or the resulting photometric amplitudes are too small for detection in the employed data.

To tackle this question, we investigated possible sets of spot parameters that satisfy the observations, i.e. those resulting in very low-amplitude variability. It is important to note that from the derived upper limits, it is not possible to derive limits for spot sizes straightforwardly. The spot-induced amplitude for a given wavelength range (or filter) depends not only on the size, contrast, and distribution of the spots, but also on the involved elements \citep{Krtic13,Prvak15}. Furthermore, the abundance patterns of BHB stars are very much different from CP stars \citep{Kinm00} and no specialized models exist to synthesize light curves for a given surface distribution of stellar spots in these objects.

As a near approximate, we therefore had to resort to the available information for several well-described mCP stars covering the $T_\text{eff}$ range of our targets: HD 3980 \citep[$T_\text{eff}=8\,300$\,K]{Nesva12}, HD 112185 \citep[$T_\text{eff}=9\,000$\,K]{Lueft03}, and HD 124224 \citep[$T_\text{eff}=13\,000$\,K]{Krtic12}. Synthetic light curves in the filters employed by the ASAS ($V$ filter), ASAS-SN ($V$ and $g$ filters), and SuperWASP (broadband filter, 4\,000 -- 7\,000\,\AA) surveys were calculated using the available surface abundance maps, TLUSTY model atmospheres, and the SYNSPEC code.

Light curves were calculated assuming spots of different sizes and contrast. If the spot contrast is kept fixed, a 50\% reduction of the spot area reduces the photometric amplitude by a factor of four in all models. On the other hand, if the area covered by spots is kept constant, a contrast reduction of 80\% is needed to achieve an amplitude change of the same magnitude. Assuming different amplitudes and keeping the spot contrast fixed, we find a clear correlation between the spot area and the resulting photometric amplitude, in the sense that spot size decreases linearly with light curve amplitude. Thus, if spots are present in our sample stars, either their contrast is much lower or, much more likely, the total area covered by spots is much smaller. This suggests that the surface characteristics of BHB and mCP stars are very different. However, this finding has been based on small number statistics and needs to be confirmed by the investigation of a larger sample of stars.

In summary, we conclude that either no stellar surface spots are present in our sample stars, or the resulting photometric amplitudes are very small and below the detection limit of our photometric data.

\section{Conclusions and outlook} \label{conclusion}

We have searched for rotationally induced variability within 30 well-established bright FHB stars. We were not able detect statistically significant photometric variability in our target stars. Either no stellar spots are present or their characteristics (contrast, total area, and involved elements) are not sufficient to produce amplitudes larger than a few millimagnitudes in the optical wavelength region. Up to now, no corresponding models exist that reliably predict photometric amplitudes for the variety of surface configurations and abundance patterns observed in BHB (and mCP) stars.

The next logical step is to search for corresponding variations in highly accurate space-based data. Unfortunately, no data for our targets are present in the CoRoT and \textit{Kepler} databases. Another important step will be the collection of highly accurate spectropolarimetric measurements of all bright FHB stars in the solar neighbourhood to search for the presence of stable magnetic fields, which will also be important for the modelling of synthetic light curves. In consequence, the combination of observational and theoretical efforts will shed more light on whether diffusion on the blue horizontal branch is able to produce spotted surfaces as has been shown for main sequence stars.

\section*{Acknowledgments}
This project was supported by the grants 7AMB17AT030 (M\v{S}MT), 16-01116S (GA~\v{C}R), and 
No. 14--50--00043 (Russian Science Foundation).
This paper makes use of data from the DR1 of the WASP data \citep{Butte10} as provided by the WASP consortium, 
and the computing and storage facilities at the CERIT Scientific Cloud, reg. no. CZ.1.05/3.2.00/08.0144, 
which is operated by Masaryk University, Czech Republic.

\end{document}